\documentclass[conference]{IEEEtran}
\IEEEoverridecommandlockouts
\usepackage{cite}
\usepackage{amsmath,amssymb,amsfonts}
\usepackage{algorithmic}
\usepackage{graphicx}
\usepackage{textcomp}
\usepackage{xcolor}
\usepackage{url}
\usepackage{subfigure}
\usepackage[misc]{ifsym}

\def\BibTeX{{\rm B\kern-.05em{\sc i\kern-.025em b}\kern-.08em
    T\kern-.1667em\lower.7ex\hbox{E}\kern-.125emX}}
\begin{document}

\title{{Proximity operations of CubeSats via sensor fusion of ultra-wideband range measurements with rate gyroscopes, accelerometers and monocular vision}
\thanks{This work is supported by the Air Force Office of Scientific Research (AFOSR), as a part of the
SURI on OSAM project “Breaking the Launch Once Use Once Paradigm” (Grant No: FA9550-
22-1-0093).}
}

\author{Deep Parikh,\thanks{Authors are with Department of Aerospace Engineering, Texas A\&M University, College Station, TX. \{deep, alikhowaja, ravikt\}@tamu.edu}
 Ali Hasnain Khowaja, Ravi Kumar Thakur, Manoranjan Majji}

\newcommand{\deep}[1]{{\color{green}Deep: {#1}}}
\newcommand{\ali}[1]{{\color{orange}Ali: {#1}}}
\newcommand{\ravi}[1]{{\color{magenta}Ravi: {#1}}}

\maketitle

\begin{abstract}
A robust pose estimation algorithm based on an extended Kalman filter using measurements from accelerometers, rate gyroscopes, monocular vision and ultra-wideband radar is presented. The sensor fusion and pose estimation algorithm incorporates Mahalonobis distance-based outlier rejection and under-weighting of measurements for robust filter performance in the case of sudden range measurements led by the absence of measurements due to range limitations of radar transceivers. The estimator is further validated through an experimental analysis using low-cost radar, IMU and camera sensors. The pose estimate is utilized to perform proximity operations and docking of Transforming Proximity Operations and Docking Service (TPODS) satellite modules with a fixed target.
\end{abstract}


\section{Introduction}
On-orbit satellite servicing has been extensively studied for refurbishing and refueling satellites, construction of large structures in space, and orbital debris management \cite{osam_godd},\cite{king2001space}. Such studies have also been extended to examine the technical and economic feasibility of robotic servicing in GEO \cite{GEO_servicing}. Most of the initial missions to service and repair satellites leverage robotic manipulators to grasp the target spacecraft\cite{VISENTIN199845}. One of the earliest spaceflight missions utilized a similar robotic manipulator to demonstrate the satellite servicing capabilities in low Earth orbit\cite{ogilvie2008autonomous}. 

However, the additional degrees of freedom associated with the floating base of the robot manipulator renders the analysis of system dynamics challenging\cite{kin_ft_base}. Furthermore, if the contact and capture of the target satellite are not planned carefully, the resulting momentum transfer after the capture can cause severe stability issues\cite{1507185}. One prominent application of the technologies developed for satellite servicing is in the mitigation of orbital debris. The cost associated with maintaining and tracking a catalog of space objects, assessing collision risk, and planning \& execution of collision avoidance maneuvers is a matter of concern in the space community\cite{SCHAUB201566}. However, due to the uncontrolled tumbling motion of the resident space objects (RSOs), robotic manipulators are often not suitable or require meticulous control strategies for proximity operations\cite{1642292}. Some of the alternatives to circumvent these limitations include the use of nets and flexible tethers \cite{BENVENUTO201645},\cite{GOLEBIOWSKI2016229} as well as harpoons\cite{DUDZIAK2015509}. Since these approaches always carry an inherent risk of unintentional damage and the breakup of the RSO, a few non-contact-based debris mitigation techniques are also being studied e.g. electrostatic disposal maneuvers\cite{SCHAUB2014110}.

Furthermore, the fast-paced advancements in the CubeSat technology has carved new frontiers in low-cost, low-risk space missions\cite{GAMBLE2014226}. The new form-factor of nano-satellites has been aptly leveraged for earth observation\cite{SELVA201250}, oceanography\cite{GUERRA2016404} and space weather monitoring\cite{https://doi.org/10.1029/2021SW003031} applications. More recently, some of the complex missions involving satellite swarming\cite{doi:10.2514/1.A35598} and rendezvous\cite{ROSCOE2018410} have also seen involvement of CubeSats. However, the form factor, limited on-board power generation, and constrained fuel storage for CubeSats have proven to be a significant obstacle in achieving reliable and robust autonomy during proximity operations and docking applications\cite{spiegel2023cubesat}. 

The Land, Air and Space Robotics(LASR) laboratory at Texas A\&M University is harnessing CubeSat technology to  detumble space objects\cite{TPODS_detumble}. The critical technology advancement is embodied in the TPODS modules. The TPODS design and operation has been demonstrated using a series of experiments\cite{TPODS_system}. One of the crucial phases of a de-tumbling maneuver using TPODS modules is the proximity operation of TPODS modules to relocate them at the preferred location on the RSO after their deployment. In addition to this, once the RSO is put into a stable attitude, it is planned to create a scaffolding structure utilizing the TPODS attached to the RSO to allow the servicing vehicle to dock with the RSO for further servicing\cite{TPODS_detumble}. Accurate pose estimation of TPODS modules is vital for proximity operations, de-tumbling and scaffolding generation processes.   

This paper presents a relative pose estimation algorithm for TPODS via fusion of UWB radar, accelerometer, rate gyro and monocular vision sensors. Section \ref{sys_d} discusses the system dynamics, including the actuator model and frictional forces associated with the devised experiment. The following section presents the sensor models for the IMU, UWB radar sensors and the camera sensor. The paper further presents the proposed estimation algorithm along with the prediction and measurement steps of the filter in Section \ref{sec_pose}. Finally, Section \ref{sec_exp} presents a selection of hardware components, experimental design for the TPODS docking experiment and the performance of the pose estimator. 

\begin{figure}[!th]
\centerline{\includegraphics[width=0.4\textwidth]{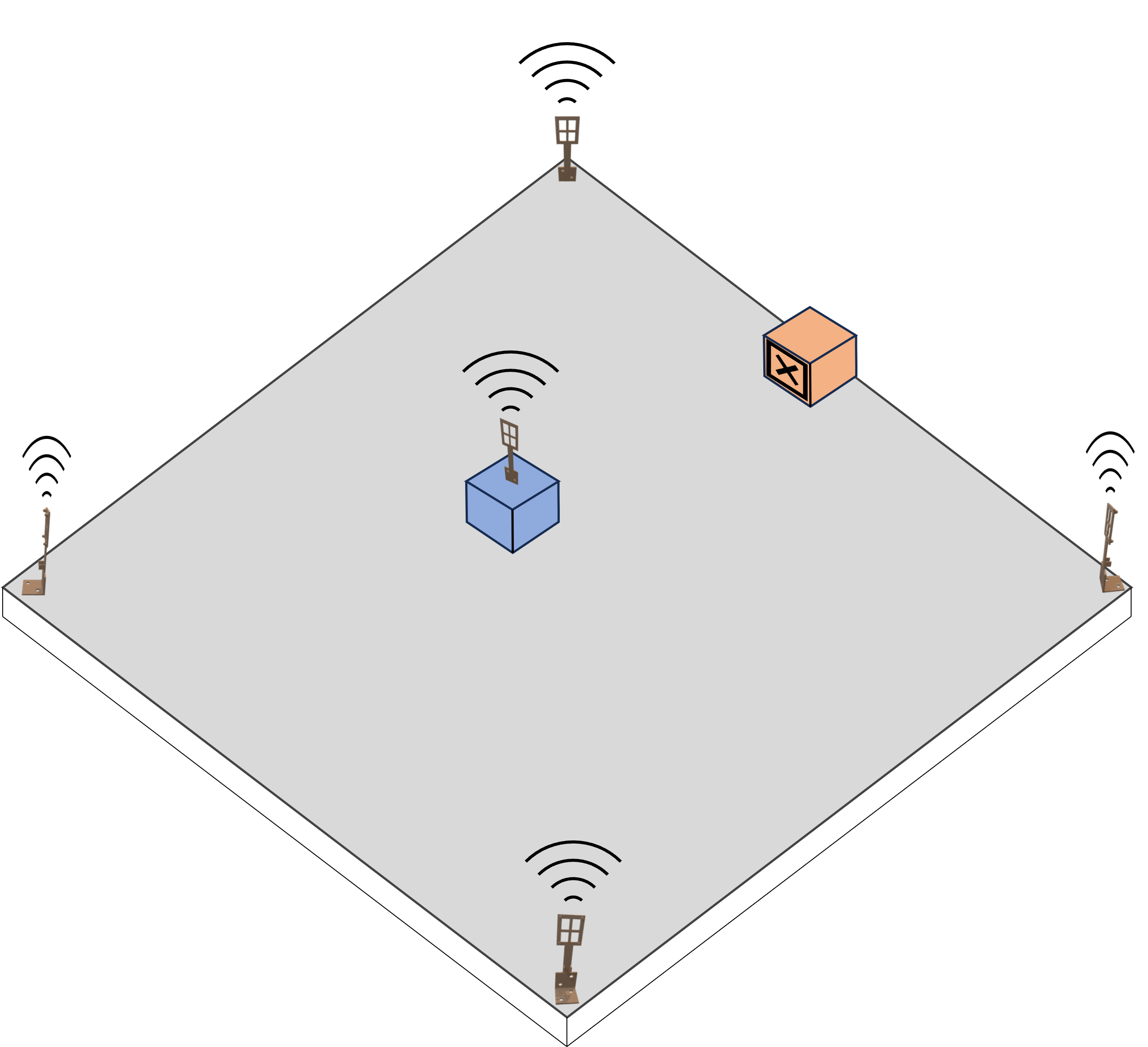}}
\caption{TPODS module uses UWB radar in two way ranging mode to measure distance to stationary anchors. Rate gyroscope measures angular velocity and accelerometer measurements enable visibility of translation velocity through indirect measurement of frictional forces. Monocular vision data is further utilized when UWB measurements are not reliable near 10 cm of the target.}
\label{fig:exp_TPODS}
\end{figure}

\section{System Dynamics\label{sys_d}}
The motion governing equations for the TPODS module are presented in this section. These equations will be refined and used in the pose estimator formulated in the later section, particularly in the prediction step. A TPODS module is considered moving in a plane, as depicted in Figure \ref{fig:exp_TPODS} and assumed to be rigid. The motion of TPODS is then governed by Newton-Euler equations
\begin{equation}
m\boldsymbol{\Ddot{x}} = \mathbf{R}[f_x\hat{e_1}+f_y\hat{e_2}+\boldsymbol{f_f}] \label{eq}
\end{equation}
\begin{equation}
\mathbf{\dot{R}} = \mathbf{R}[\![\boldsymbol{\omega}\times]\!]
\end{equation}

 Where the position of TPODS module in the plane of motion with respect to inertial frame is denoted by $\boldsymbol{x}$ and the orientation of the body with respect to the inertial frame is given by $\boldsymbol{R}$. Total external force exerted by the on-board thrusters in body-fixed $x$ and $y$ axis is given by $f_x$ and $f_y$ respectively. The total frictional forces in the body-fixed reference frame is represented by $\boldsymbol{f_f}$. Further, $\boldsymbol{\omega}=\omega_3\hat{e_3}$ is the angular velocity of the module expressed in the body frame and $[\![\boldsymbol{\omega}\times]\!]$ represents the standard cross product. 

\subsection{Actuator thrust force}
TPODS module has four nozzles mounted the in `X' configuration and release precise jet of high pressure air stored in the accumulator \cite{TPODS_system}. A comprehensive system identification exercise has been performed to determine an accurate model of each thruster\cite{TPODS_estm}. Based on the applied input duty cycle and number of nozzle active, values of $f_x$ and $f_y$ are computed at each time instance and used in the state propagation step of Section \ref{sec_pose}.

\subsection{Frictional forces}
For the TPODS docking experiment workbench, each modules experience significant frictional forces from the contact between the spherical ball transfers and the acrylic sheet. The frictional force is a function of the loading of the ball transfers and the local velocity\cite{ball_transfer}. For the analysis presented in this paper, the frictional force is considered to be related to the local velocity of the module via proportionality constant $\mu$, and its value is computed based on the loading of each ball transfer. It is also important to note that due to the physical symmetry of three ball transfers, the effect of friction due to translation and rotation can be considered independent. It is assumed that the TPODS moves without any slip and all ball transfers share equal load.
\begin{equation}
\boldsymbol{f_f} = \mu m g \mathbf{R}^{-1}\boldsymbol{\dot{x}}\label{eq_ff}
\end{equation}

\section{Sensor models and Selection\label{sen_mod}}
The primary sensors considered for this study are rate gyroscope, accelerometer, UWB radar and monocular vision. The sensor models for rate gyroscope, accelerometer and UWB radar are discussed in \cite{OG_fusion} and adopted in this paper. The following subsections discuss the sensor selection criteria for the experimental analysis and measurement model for the monocular vision sensor.

\subsection{Inertial Measurement Unit}
The typical inertial measurement unit consists of accelerometers, gyroscopes and magnetometer sensors. Each sensor provides measurements of different physical quantities and their behavior are studied in ample existing literature \cite{IMU_review}. For brevity, such analysis is not repeated in this paper. However it is the preliminary driving factor for selecting the hardware employed during the experimental study. Since the measurements from these sensors contain several deficiencies, involved signal processing and estimation algorithms are employed to compute precise attitude. These algorithms are computationally expensive and often require accurate information about sensor characteristics. Hence, the VectorNav VN-100 AHRS sensor is
leveraged for the TPODS docking experiment to avoid the additional complexity associated with the attitude estimation from IMU. The VN-100 consist of an internal compute unit and a processing engine, along with factory calibration parameters, to ensure precise attitude estimates. 

Accurate attitude estimates and their respective angular velocities are assumed to be available with their respective measurement noise. In addition, filtered acceleration data is also made available from VN-100. The acceleration measurement are leveraged to infer the frictional forces considering the following system dynamics
\begin{equation}
\boldsymbol{z}_{acc} = \mathbf{R}^{-1}\boldsymbol{\Ddot{x}}+\boldsymbol{\eta}_{acc} =\frac{1}{m} \left(f_x\hat{e_1}+f_y\hat{e_2}\right)+\mu g \mathbf{R}^{-1}\boldsymbol{\dot{x}}\label{eq_acc}
\end{equation}
The acceleration measurement and their association with frictional forces render velocity of the TPODS module observable due to the dependency of the frictional forces on the velocity. Here, the term $\boldsymbol{\eta}_{acc}$ represents the zero-mean white noise. 

\subsection{Monocular Vision sensor}

A monocular camera is chosen to maintain smaller size, weight, and power for the TPODS modules as opposed to stereo-vision. The monocular camera can be modeled with the pinhole projection model\cite{Xu1996EpipolarGI} which assumes a single point camera aperture with no lenses focusing the light. Mathematically this model is described as
 \begin{eqnarray}
 u_i' =  \frac{u_i-u_o}{f_{cx}} = \frac{X_i^C}{Z_i^C} \label{eq:up} \\
 v_i' =  \frac{v_i-v_o}{f_{cy}} = \frac{Y_i^C}{Z_i^C} \label{eq:vp}
 \end{eqnarray}
where ($u_o,v_o$) is the point at which the optical axis intersects the image plane, ($u_i$,$v_i$) is the image plane projection of the $i^{th}$ feature, ($u_i'$,$v_i'$) is the rectified image projection of the $i^{th}$ feature, ($X_i^C$,$Y_i^C$,$Z_i^C$) is the 3D coordinate location of the ith feature expressed in the camera frame, and $f_{cx}$ , $f_{cy}$ are the focal lengths between the image plane and camera frame. Camera measurements are affected by a zero-mean Gaussian white noise $\nu_{ui}$ and $\nu_{vi}$ which results in the following camera sensor model
\begin{eqnarray}
    \Tilde{u_i}' = u_i' + \nu_{ui} \\
    \Tilde{v_i}' = v_i' + \nu_{vi} 
\end{eqnarray}
The chosen monocular vision sensor is the OpenMV H7 Plus Camera. The OpenMV Cam is a small, low power, encapsulating a microcontroller board with its own machine vision libraries to be leveraged in numerous applications. 

\subsection{UWB range measurements}
The UWB radio modules mounted on four corners of the workbench, as well as on the TPODS module, utilize the Time Of Arrival(TOA) based two-way ranging(TWR) algorithm to determine the relative distance between TPODS and each fixed radar module. It is known that the clock drifts between radio modules and occasional outliers in the TOF measurements render the measured distance unreliable. Similar to the IMU measurements, several filtering techniques must be administered to ensure reliable distance measurements. Honoring the design philosophy of distributed computing, the Loco Positioning System from Bitcraze AB is selected. More details on the implementation and the hardware setup are provided in Section \ref{sec_exp}.   

\section{Pose Estimator\label{sec_pose}}
A discrete extended Kalman filter(EKF) based estimator is leveraged to estimate the state of the TPODS module. The discrete EKF formulation of \cite{crassidis2011optimal} is adopted for this paper and the equations are not repeated here for the sake of brevity. The goal of the estimator is to combine the measurements from the rate gyroscope, accelerometer, UWB radar and monocular vision along with the knowledge about system dynamics presented in Section \ref{sys_d}, sensor models of Section \ref{sen_mod} and the current control input to compute the best guess of the current inertial position and velocity of the TPODS module.

A caret sign represents an estimated value of each quantity. It is assumed that accurate measurements for the angular velocity $\omega$ and the orientation $\psi$ are available, with the respective measurement noise. Hence, 
\begin{equation}
\boldsymbol{\hat{\omega}} = \boldsymbol{z}_{gyro}
\end{equation}
\begin{equation}
\hat{\psi} = z_{\psi}
\end{equation}
In addition to this, the estimator also predicts a stochastic state vector,
\begin{equation}
\boldsymbol{\zeta} = (\boldsymbol{x},\boldsymbol{\rho})
\end{equation}
Where $\boldsymbol{\rho} = \mathbf{R}^{-1}\boldsymbol{\dot{x}}$ is the body velocity of the TPODS module. 
\subsection{State Prediction}
Starting with an initial guess of the mean of the states, the system dynamics given by Equations \ref{eq} and \ref{eq_ff} are utilized to predict the mean of states at the next time instance. Similarly, the system dynamics Jacobin is computed and further combined with the current state covariance and process noise to predict the state covariance for the next time instance\cite{crassidis2011optimal}.

\subsection{Rate gyroscope measurement update}
Since angular velocity is not an explicit state, the rate measurements from the gyroscope do not have a direct impact during the measurement update. However, the state propagation requires knowledge of angular rates and current orientation. 

\subsection{Accelerometer measurement update}
Equation \ref{eq_acc} is employed to compute the predicted acceleration measurements. Here, the true quantities are replaced with their respective state estimates and the current orientation measurement from the VN-100 IMU is used to compute the predicted acceleration measurements. Equation \ref{eq_acc} has a linear dependence on the body-fixed velocity of the TPODS module, which renders the velocity observable from acceleration measurements. This is leveraged to incorporate acceleration measurements in the EKF formulation where only explicit states are inertial position and velocity in the body-fixed frame. 

\subsection{Vision based pose estimation}
The vision based pose estimation employs sensor fusion of camera with IMU measurements via an EKF. The state vector 
of the filter consists of the attitude ($\textbf{q}$), position ($\boldsymbol{x}$), and velocity ($\boldsymbol{\rho}$) of the TPODS chaser module
\begin{equation}
    \boldsymbol{\zeta_v} = \left(\textbf{q},\boldsymbol{x}, \boldsymbol{\rho}\right)    
\end{equation}
The camera measurement model is given by 
\begin{eqnarray}
    u_i' = \frac{r_{11}X_i^B + r_{12}Y_i^B + r_{13}Z_i^B + x}{r_{31}X_i^B + r_{32}Y_i^B + r_{33}Z_i^B + z} \label{eq:uip}\\
    v_i' = \frac{r_{21}X_i^B + r_{22}Y_i^B + r_{23}Z_i^B + y}{r_{31}X_i^B + r_{32}Y_i^B + r_{33}Z_i^B + z}\label{eq:vip}    
\end{eqnarray}
where the $r_{ij}$ components are the rotation matrix representation of the attitude ($\textbf{q}$), $(u_i',v_i')$ is the rectified image projection of the ith feature, and ($X_i^B$,$Y_i^B$,$Z_i^B$) is the 3D coordinate of the ith feature expressed in the target TPODS module body frame. The sensitivity matrix needed for the measurement update can be calculated with
\begin{equation}
    H_i = \begin{bmatrix} \frac{\partial u_i'(\boldsymbol{\zeta_v})}{\partial\boldsymbol{\zeta_v}} \\ \frac{\partial v_i'(\boldsymbol{\zeta_v})}{\partial\boldsymbol{\zeta_v}}\end{bmatrix}
\end{equation}
where the subscript $i$ denotes the $i^{th}$ feature's sensitivity matrix. Given the initial pose estimate the same state propagation and measurement update equations previously discussed can be used to update the states along with the camera measurement model. 

\begin{figure}[b!]
\centerline{\includegraphics[width=0.4\textwidth]{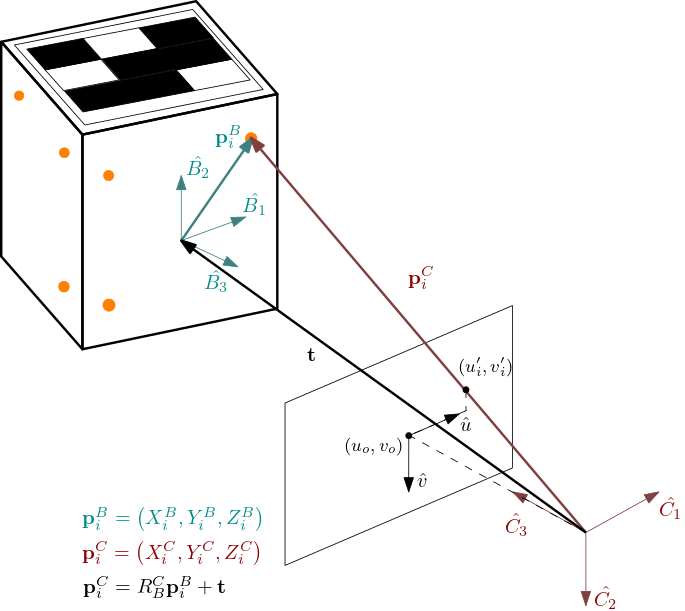}}
\caption{Geometry of the TPODS System perspective projection problem. The camera reference frame ($\Hat{C}$) which is on the chaser TPODS module can be related to the target TPODS module ($\Hat{B}$) by relating the respective position vectors to the ith feature.}\label{fig:geometryTPODS}
\end{figure}

The filter begins with an initial pose estimate given by a pose estimation algorithm. The camera sensor uses LED marker based feature for establishing 3D-2D correspondences. The three-dimensional coordinate of the markers are defined in the body-centred coordinate frame of the TPODS module and relates to camera frame as shown in Figure \ref{fig:geometryTPODS}. For perspective-n-point (PnP) projection based pose computation, a minimum of three points are required. Thus, three markers are used to form pattern on each face of the CubeSat module. In contract to a PnP approach implemented in \cite{MonoPoseEst_TPODS}, the current work employs a perspective-3-point projection algorithm \cite{kneip2011novel}.
 
The selection of the algorithm is driven by the hardware Size, Weight and Power (SWaP) constraint imposed by the camera. The method introduces intermediate frames created using the feature vector from camera and image vectors. The solution to pose problem is obtained by solving relative pose between these intermediate frames. Finally, the algorithm directly solves for rotation and translation vectors. 

Estimation of pose from three feature point is a challenging problem due to symmetry of the object, as this can result in solutions which are ambiguous. A robust pose computation can be challenging as there are only minimum number of distinctive features needed for the pose estimation. To address this issue, a face detection method based on extracting blobs\cite{kong2013generalized} is devised which leverages the geometric relation between three dimensional coordinates of LED markers. The detected face aids in identifying unambiguous feature positions for the pose estimation process. 

\begin{figure}[tbp]
\centerline{\includegraphics[width=0.5\textwidth]{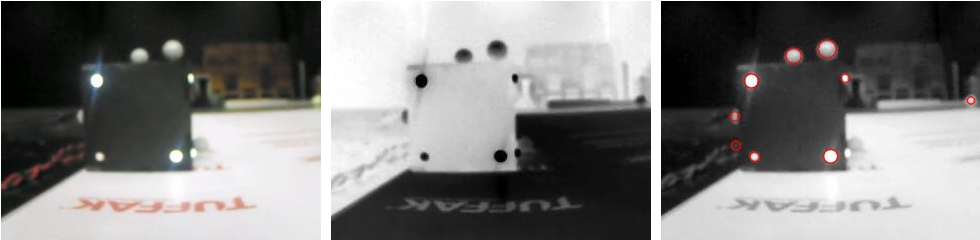}}
\caption{The TPODS module uses LED based marker patterns, which are unique for each face, for monocular vision based pose estimation. The monocular camera captures an image frame and converts it in grayscale, which is further used to obtain all the blobs present in the image. The blobs outside certain threshold are rejected and only unique features are used for pose estimation.
} 
\label{fig:range_meas}
\end{figure}

\subsection{UWB range measurements update}
A distinctive characteristic of UWB radar measurements is the presence of outliers due to clock drift, multi-path reflection and non-line-of-sight propagation between two radar modules\cite{9372785}. The default firmware of the Loco positioning node contains a few layers of preliminary error handling to tackle clock drift \footnote{\url{https://www.bitcraze.io/documentation/repository/lps-node-firmware/master/functional-areas/tdoa3_implementation/#error-management}}. While the current arrangement of UWB radar modules are effective in keeping the outliers due to multi-path reflection and non-line-of-sight propagation low, it is important to note that extension of such pose estimator algorithm for full six degrees of freedom might require robust estimation algorithms like error-state unscented Kalman filter\cite{9841385}.

Figure \ref{fig:range_meas} shows synthetic measurements for the UWB range sensor pair. Furthermore, simulated measurements for the accelerations, angular velocity and orientation are also generated for the same trajectory with respective measurement noise. The Mahalanobis distance provides a degree of statistical agreement of the measurements based on the known plant and sensor models\cite{DEMAESSCHALCK20001}. Hence, an outlier rejection scheme based on the Mahalanobis distance has been chosen to achieve robust state estimates. A squared Mahalanobis distance has been computed for each measurement at each time instance, and the individual measurements which fall outside the $99.99\%$ probability gate are rejected during the measurement update. The effectiveness of this strategy is evident in Figure \ref{fig:range_meas}, as the curve for the estimated range is reasonably smooth.

\begin{figure}[t]
\centerline{\includegraphics[width=0.5\textwidth]{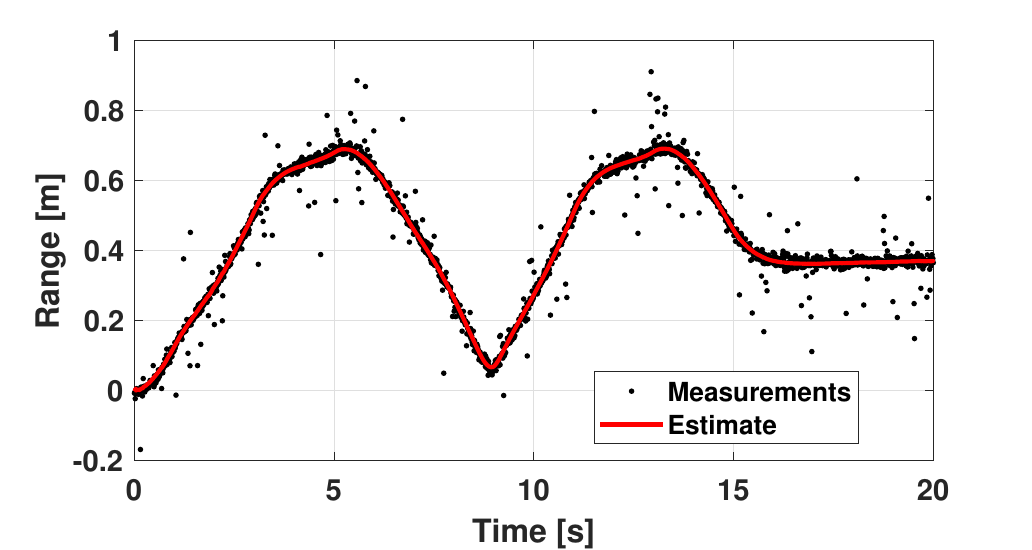}}
\caption{Simulated range measurements are corrupted with measurement noise and outliers. The module is commanded to move under constant acceleration in X and Y directions alternatively. The range measurements are sampled from a Gaussian distribution with a true range as mean and standard deviation of 1 cm. For the outlier generation, $10\%$ of the total samples are randomly chosen and re-sampled from a Gaussian distribution with a standard deviation of 10 cm.}\label{fig:range_meas}
\end{figure}

\section{Experimental Validation\label{sec_exp}}
The proposed pose estimation via sensor fusion is validated through experimental analysis at the LASR laboratory. The TPODS module MK-III\cite{TPODS_system}\cite{TPODS_estm}, which has been devised and fabricated at the LASR laboratory, is used as a chaser spacecraft and made to move under the influence of a predefined input sequence. The propulsion subsystem of the TPODS module consists of two FESTO® VUVG solenoid valves and four EXAIR® Atto Super Air Nozzles. The nozzles release a precise jet of compressed air stored inside a PET accumulator when the solenoid valve is commanded to open via a central computer. The respective reaction force on the module is leveraged to achieve controlled motion.

The module also consists VectorNav® VN-100 IMU sensor. The on-board processing engine of VN-100 runs an AHRS, which provides precise attitude, angular rate and acceleration information at a $100Hz$ update rate. Since the motion of the module is restricted in a plane, four Loco Positioning nodes are used as anchors. The module is also equipped with a similar poisoning node, albeit running a modified firmware. When configured in the TDoA3 mode, each loco node computes the relative distance to the other nodes present in the vicinity via randomized TWR. The node mounted on the module relays the range from each anchor to a central computer. 

The module also includes an OpenMv H7 monocular vision camera with an onboard processor running the vision-based pose estimation algorithm presented in earlier sections. The camera is commanded to initialize the pose estimation once in close proximity to the target. For long-range operations, only UWB radio and IMU measurements are used. 

Teensy 4.1 serves as the central computing unit for the TPODS module. As evident from the sensor selection strategy presented in Section \ref{sen_mod}, each sensor has its own compute power. Hence, the overall computational loading of the central computer is significantly reduced. The computer requests orientation, angular rate and acceleration data from the IMU at $100Hz$, UWB range data at approximately $50Hz$ and the monocular vision-based pose estimate at $20Hz$. The estimator fuses the UWB range with the IMU data to estimate the pose of the module for the long-range operations. Once within the vicinity of the target, OpenMV camera is commanded to initialize pose estimation and, after achieving sufficient confidence in the vision based pose estimate, it is treated as the primary measurement along with IMU measurements. 

The LASR laboratory is equipped with overhead motion capture system that can measure and record the module's position and orientation with an accuracy of a few millimeters and degrees respectively. The pose is logged at $120Hz$ and further used as a ground truth for the evaluation of the performance of the pose estimator presented in this paper.

\begin{figure}[b!]
\centering
\includegraphics[width=0.45\textwidth]{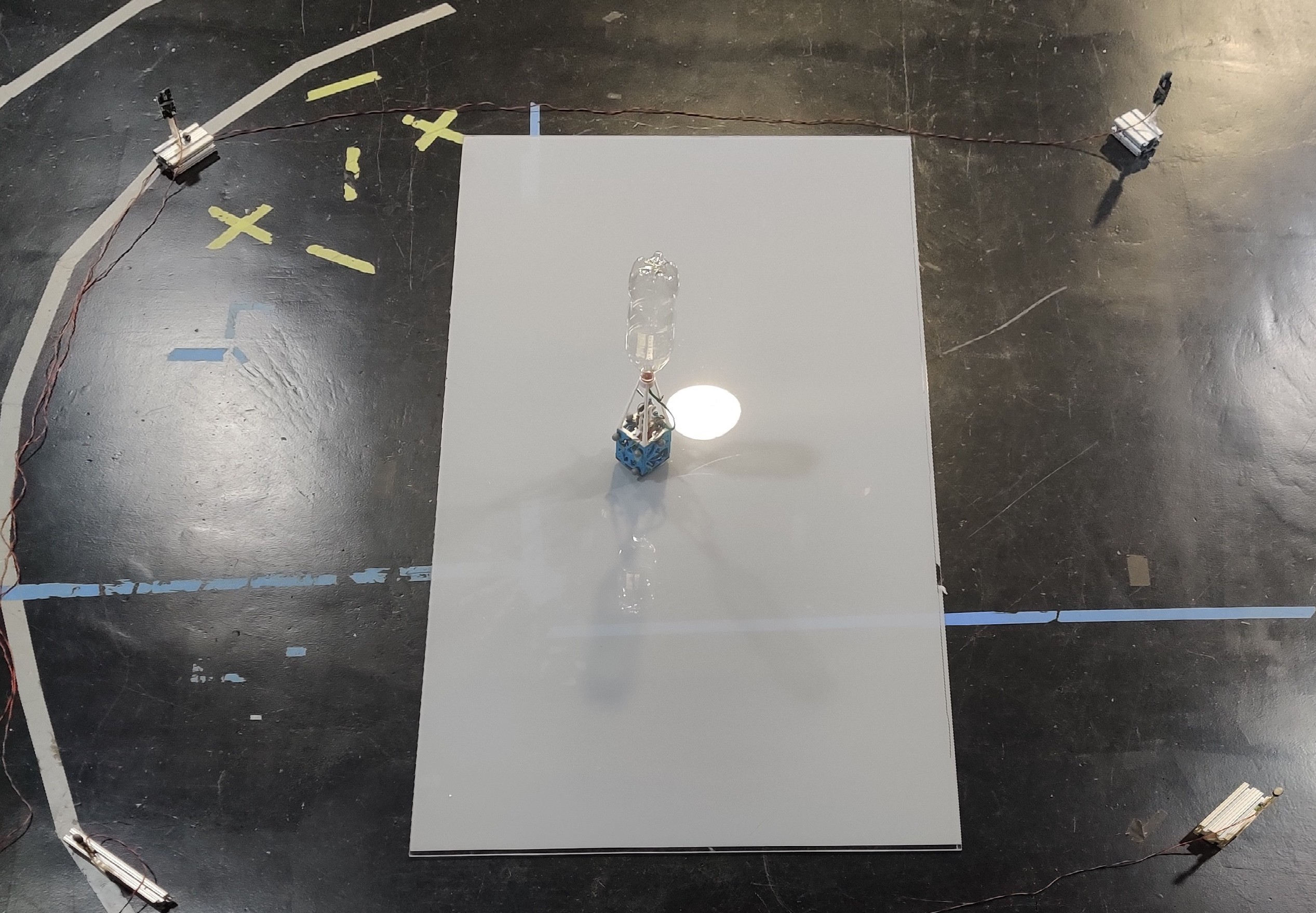}
\caption{The DWM1000 UWB radio used in the loco node possesses different gains for different planes of motion\cite{sorgel2005influence}. This is undesirable and might result in different pose estimation biases while moving in various planes \cite{OG_fusion}. However, since the module is constrained to move in a single plane, such effects are not of concern for this study. Each anchor node is mounted above $15cm$ from the ground to eliminate any interference of the multi-path reflection from the ground. In addition, the node on the module is mounted on the top to have a clear line of sight with all four anchors. The accompanying video submission includes more details about the setup.}
\label{fig:exp_results}
\end{figure}

\subsection{Estimator tuning parameters}
The module's mass is $0.795kg$ and the moment of inertial along the principal rotation axis is $4.987689e^{-3}kgm^2$. The coefficient of friction is chosen as $\mu=0.015$, considering the operating velocity and the loading of ball transfers. 

For the EKF, the measurement noise covariance of the accelerometer is selected as $\eta_{acc}=diag(0.05g,0.05g)^2m^2s^{-4}$ and the noise covariance of the UWB range measurements is taken as $\eta_{uwb}=(0.08)^2m^2$. These values are partly from the manufacturer data sheet and partly from the iterative tuning performed on the collected data set from the experimental setup. Similarly, the noise covariance of the monocular pose measurements is computed as $\eta_{vision}=diag(0.001m^2,0.001m^2,0.00225rad^2)$ from comprehensive experimental analysis. For the outlier rejection, a squared Mahalanobis distance of 15.14 is chosen as the threshold.

In addition, the process noise of covariance $diag(8E^{-4},8E^{-4})m^2s^{-4}$ is chosen to act on acceleration. This ensures that the covariance of the pose estimate stays inflated as the velocity states explicitly affect the inertial position.

\subsection{Experiments and Performance Analysis}
The verification of the pose estimation algorithm is done via two different experimental scenarios, motion towards a predefined target location, only using the UWB radar and IMU measurements, and docking with a fixed target. Figure \ref{fig:exp_results} shows a qualitative performance of the pose estimator for the first experiment and it is also included in the accompanied video. The numerical performance is inferred by computing the spared of the separation distance of the module from the desired target location at the end of the first experiment.

$1)$ \textit{Closed-loop Position Control} : 
The TPODS module is commanded to reach a perdefined target location in the inertial frame from an arbitary starting position and orientation. Sufficient range sample has been processed at the beginning to gain confidence for the position estimate. Once the covariance of the position estimate settles to a value below predetermined threshold, the module first establishes a LOS with the target by correcting the orientation. The next phase is classical LOS guidance where the objective is to reduce the separation distance from the target. It is important to note that only UWB radar and IMU measurements were used for the pose estimation during this experiment.

\begin{figure}[t!]
\includegraphics[width=0.45\textwidth]{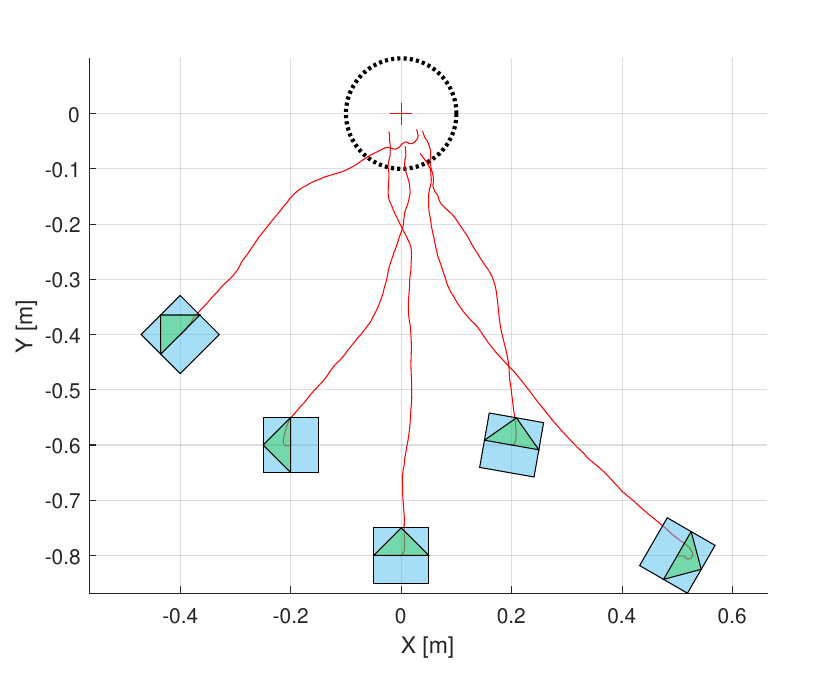}
\caption{Experimental validation of the pose estimator for waypoint navigation. The module is commanded to reach a predefined position marked with red '+'. It is evident that based on the initial position and orientation, the module takes different trajectories to reach the target location. Based on the spread of the end-point, it can be concluded that the pose estimator is accurate within $\pm 5cm$ of the true position. This aligns well with the strategy of shifting to vision based pose estimation once within $10cm$ of the target.}
\label{fig:exp_results}
\end{figure}

\begin{figure}[t]
\centerline{\includegraphics[width=0.4\textwidth]{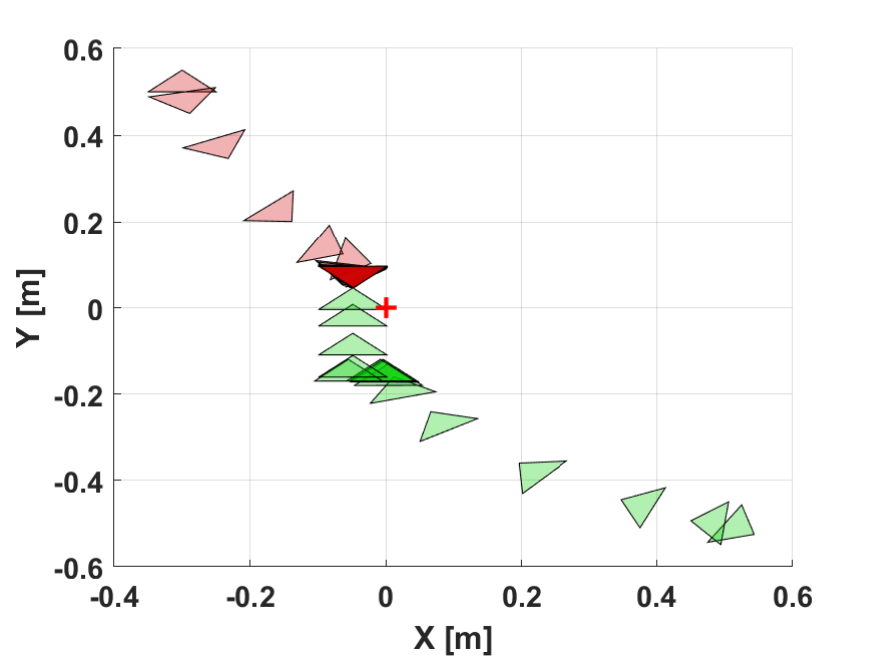}}
\caption{Guidance logic for TPODS docking experiment consists of LOS guidance to a predefined position for long-range operations, attitude alignment when both modules are in close proximity, and deputy-chaser proximity operations once the vision-based pose estimation is reliable. Each TPODS
module is represented with a colored triangle with the triangle head indicating the module's orientation at various simulation instances. The darker shade depicts that the module spent longer at a particular position. TPODS module represented with red triangle assumes the role of deputy in this simulation.}
\label{fig:GNC_TPODS}
\end{figure}

$2)$ \textit{Docking with fixed target} :
Once the performance of the pose estimator with sensor fusion of UWB radar and IMU was verified, the integrated docking experiment was conducted to validate the intended application of this exercise. The TPODS module is commanded using a combination of LOS and terminal guidance, as shown in Figure \ref{fig:GNC_TPODS}. While the simulations were performed considering both TPODS mobile, the experiment was conducted with one of the TPODS modules fixed. The fixed target module only contains LED markers and magnetic attachment points. The chaser TPODS starts at a random orientation without LOS to the target. The LOS guidance commands the module first to achieve the LOS with the target and then close in the distance to the target while maintaining the LOS. Once within $10cm$ of the target, the guidance logic is switched to terminal guidance, first step of which is to achieve position lock on the target using monocular vision, followed by directly facing the target via removing any cross-track errors, and finally closing in towards the attachment point.

\section{Conclusion}
A robust, low-cost pose estimation via fusion of UWB radar modules, IMU and monocular vision sensors has been proposed and validated in this paper. It has been shown that the pose estimation methods are suitable for proximity operations of TPODS. The limitation of unreliable range measurements from UWB radio while operating within $10cm$ of the target is addressed by adding a monocular vision-based pose estimator. The UWB radio and monocular vision provide a conducive complement for range measurement and the accelerometer and rate gyroscope aid the pose estimation process. Consequently, adaptation of the methods proposed in the paper are anticipated to provide a reliable method to navigation multiple vehicles carrying out proximity operations in space. However, extension of this approach for non-planar motion might suffer from well-known error sources such as multi-path reflection and non-line-of-sight propagation. 

\section*{Acknowledgment}

Program monitors for the AFOSR SURI on OSAM, Dr. Frederick Leve of AFOSR, and Mr. Matthew Cleal of AFRL are gratefully acknowledged for their watchful guidance. Prof. Howie Choset of CMU, Mr. Andy Kwas of Northrop Grumman Space Systems and Prof. Rafael Fierro of UNM are acknowledged for their motivation, technical support, and discussions. Mr. Kristoffer Richardsson from Bitcraze AB is also acknowledged for their insightful interactions on the Loco Positioning system. 





\bibliographystyle{ieeetr}
\bibliography{./bib/refsm-astro}

\end{document}